# Modulation of mid-IR radiation by a gated graphene on ferroelectric substrate

M.V.Strikha.

*V.Lashkariov Institute of semiconductor physics, NAS of Ukraine*
*Nauky ave., 45, Kyiv-28, 03650, Ukraine*
*maksym_strikha@hotmail.com*

SUMMARY. The possibility of modulation of mid-IR radiation by low voltage gated single/multilayer graphene on $Pb(Zr_xTi_{1-x})O_3$ (PZT) ferroelectric substrate was demonstrated theoretically. The depth of modulation for 5 layer graphene for the range of gate voltages, where thin PZT epitaxial film behaves as high-κ dielectric, can be of 10% order.

## 1. Introduction

A graphene-based optical modulator for near-IR range (1.35-1.6 μm) was realized in [1]. It was demonstrated, that such a modulator can be perspective for the devices with on-chip optical interconnections, and it's modulation efficiency is already comparable to, if not better than, traditional semiconductor materials such as Si, GeSi, InGaAs, which are orders of magnitude larger in active volume. In this work single layer graphene was placed on 7 nm thick $Al_2O_3$ substrate over the Si gate, which had also served as a waveguide for near-IR radiation; the device footprint was as small as 25 $\mu m^2$, and the operation speed as high as 1.2 GHz.

The general theory of the carrier induced modulation of radiation by a gated graphene has been worked out in [2]. It was demonstrated, that the carriers contribution modifies essentially the graphene response due to the Pauli blocking effect, when absorption is suppressed at $\hbar\omega/2 < E_F$, where $E_F$ is Fermi energy. At low temperatures, or at high doping levels, the threshold frequency for the jump of absorption (when absorption becomes essential) is determined by condition:

$$\hbar\omega_{th} = 2E_F \sim \sqrt{n}, \qquad (1)$$

where concentration $n$ depends linearly on gate voltage $V_g$, substrate permittivity κ, and is inverse to substrate width $d$:

$$n(cm^{-2}) = 7.2 \cdot 10^{10} (\frac{300}{d(nm)})(\frac{\kappa}{3.9})V_g(V) \qquad (2)$$

The figures in (2) are normalized for the characteristics of the SiO$_2$ 300 nm thick substrate, used in first graphene-related work [3]. However, the use of high-κ substrates (AlN, Al$_2$O$_3$, HfO$_2$, ZrO$_2$) enables to obtain higher concentrations for the same gate voltages (see e.g. [4]). This is important, because for the threshold wavelength $\lambda_{th}$ corresponding threshold frequency (1), eqs. (1, 2) yields:

$$\lambda_{th} \sim \sqrt{\frac{d}{V_g}} \equiv 1/\sqrt{E_s} \qquad (3)$$

Here $E_s$ - homogeneous gate-induced electric field intensity in the substrate. Eq.(3) yields, that in order to get modulation for the smaller wavelength of radiation (i.e. in visible range) one needs stronger fields (and higher gate voltages), which finally can cause the breakdown of the substrate.

The calculations, performed in [2], demonstrate, that the modulator for telecommunication near-IR range (~1.5 µm) can be based on single-layer/multilayer graphene, placed over high-κ substrate (in fact, this was the case, realized in [1], see Fig.1). The effective modulation can be realized in this case for the applied fields ~ 5 MV/cm. For the case of low-κ SiO$_2$ substrate the field should be essentially stronger ~ 20 MV/cm, which is comparable to breakdown value. The highest values of $E_s$, reached in [1], were in fact of 5 MV/cm order. This needed, however, the extremely high accuracy of substrate preparation (7 nm thick Al$_2$O$_3$ substrate was deposited over the Si gate, which had also served as a waveguide, by atom layer deposition).

## 2. Mid-IR modulation

The use of ferroelectric substrates with extremely high permittivity can be fruitful for the further development of the gated graphene based modulators. The study of graphene on such a substrate was carried intensively in recent time (see e.g. [5-7]). It was demonstrated there, that an unusual resistance hysteresis occurs in gate sweeps at high voltages due to substrate ferroelectric properties. However, at low voltages ($V < V_{cr} \sim$ 1-2 V) the epitaxial ferroelectric Pb(Zr$_x$Ti$_{1-x}$)O$_3$ (PZT) thin films behave as a high κ dielectric with κ = 73 (x = 0.2 [5]), κ = 400 (x = 0.3 [6]). This permits to use them in low voltage mid-IR gated graphene based modulators.

The critical value for the field in PZT substrate, under which this substrate still behaves as a high κ dielectric, can be obtained from [5] ($d$ = 300 nm, $V_{cr} \sim$ 2

V) and [6] ($d$ = 360 nm, $V_{cr}$ ~ 1 V). This yields $E_{cr}$ = 67 kV/cm (x = 0.2), $E_{cr}$ = 28 kV/cm (x = 0.3).

Fig.2 presents the threshold wavelength (3) dependence on the field in the substrate for the different substrates: $SiO_2$ (curve 1), $Al_2O_3$ (2), $ZrO_2$ (3), PZT (x = 0.2, 4), PZT (x = 0.3, 5).

As one can see from fig.2, the fields, corresponding $\lambda_{th}$ for telecommunication range ($\lambda$ = 1.55 μm, horizontal line), are several times higher for PZT, than $E_{cr}$. However, for mid-IR range (namely for $\lambda$ = 10.6 μm, corresponding the $CO_2$ laser wavelength) modulation can be realized for the fields essentially lower than $E_{cr}$, when PZT behaves as high κ dielectric with extremely high permittivity.

The transition and reflection coefficients of the graphene layer – PZT substrate – Si gate system are introduced as (see [2]):

$$T_\lambda = \sqrt{\kappa_{Si}(\lambda)} \frac{|E_t|^2}{E_{in}^2}; R_\lambda = \frac{|E_r|^2}{E_{in}^2} \qquad (4)$$

Here κ is Si gate permittivity, dependent on wavelength. The correlation between the amplitudes $E$ of the incident (*in*), reflected back into vacuum (*r*), and transmitted into Si gate (*t*) waves can be obtained similarly to it was done in [2] by solving the system of wave equations in vacuum, substrate, and gate, with the proper boundary conditions, taking into consideration the absorption due to interband carriers transitions in graphene layer. These coefficients, calculated for single layer (a) and 5-layer graphene (b) on PZT substrates with x = 0.3 for the films of different thickness (240 nm, 280 nm, 320 nm, 360 nm, 400 nm), are presented in fig.3,4. The dielectric permittivity of PZT at $\lambda$ = 10.6 μm is taken as 5 (see [8]).

As one can see, the jump in absorption at approximately 2.5-3 kV/cm for single layer graphene, and at 13-15 kV/cm for 5 layer graphene on PZT with x = 0.3 (where permittivity is several times higher than in PZT with x = 0.2) leads to essential jump in reflection and transmittivity. The deepness of modulation can be of 20% order for the fields, much lower than the critical ones, at which the ferroelectric hysteresis phenomena start.

## 3. Concluding remarks.

The results obtained demonstrate the possibility of the fabrication of low voltage gated graphene on PZT ferroelectric substrate based modulator for mid-IR range. The advantage of such a modulator in comparison with one realized in [1] with atomically deposited 7 nm thick $Al_2O_3$ substrate can be a comparative simplicity of the epitaxial PZT films substrates preparation. This modulator can potentially operate at 500 GHz, because the times of speed limiting processes of carries recombination and generation in graphene are of picoseconds order.

The modulation of near-IR can be based on the same mechanism, but for the substrates of PZT with higher κ. It is known, that the PZT features an extremely large κ at the morphotropic phase boundary near $x$ = 0.52 [9]. The dielectric constant of PZT can range to 3850 depending upon orientation and doping. The authors of [7] had observed κ = 2000 by substitutional doping of Pb by La and by fine tuning the ratio between Zr and Ti. Such a value can be quite sufficient for near-IR modulation.

The results obtained should stimulate the experimental study of the electrooptical modulation of near- and mid-IR radiation by structure of single layer/multilayer graphene on ferroelectric substrate.

**Acknowledgment**

This work was supported by State Fundamental Research Fund of Ukraine (Grant 40.2/069).

**Captions for figures**

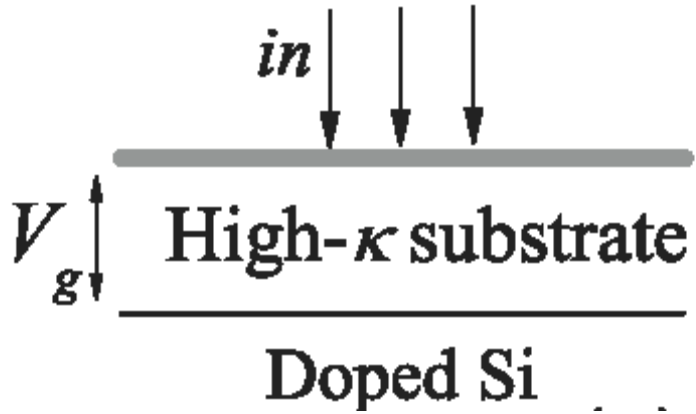

Fig.1. Single-layer/multilayer graphene, placed over high-κ substrate on Si gate ($V_g$ is applied gate voltage).

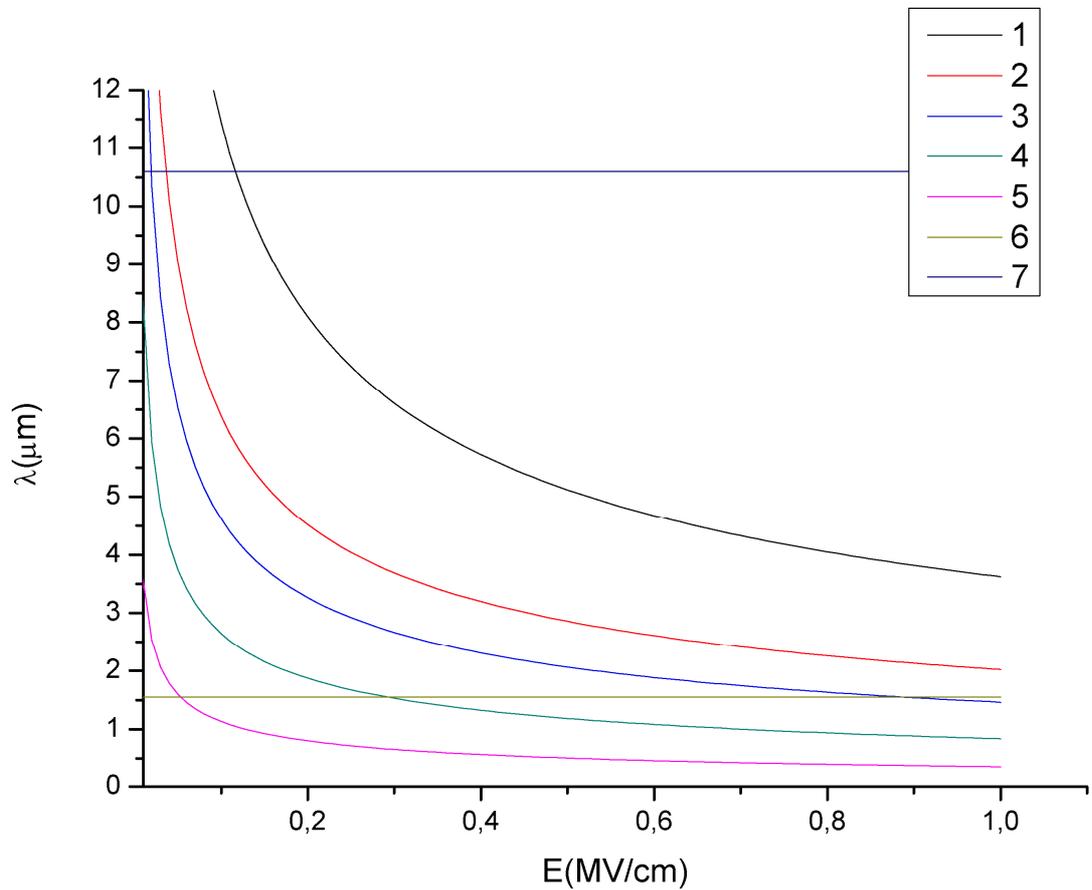

Fig.2. Threshold wavelength dependence on the field in the substrate for the different substrates: $SiO_2$ (curve 1), $Al_2O_3$ (2), $ZrO_2$ (3), PZT (x = 0.2, 4), PZT (x = 0.3, 5). Lines 6 and 7 correspond 1.55 and 10.6 μm respectively.

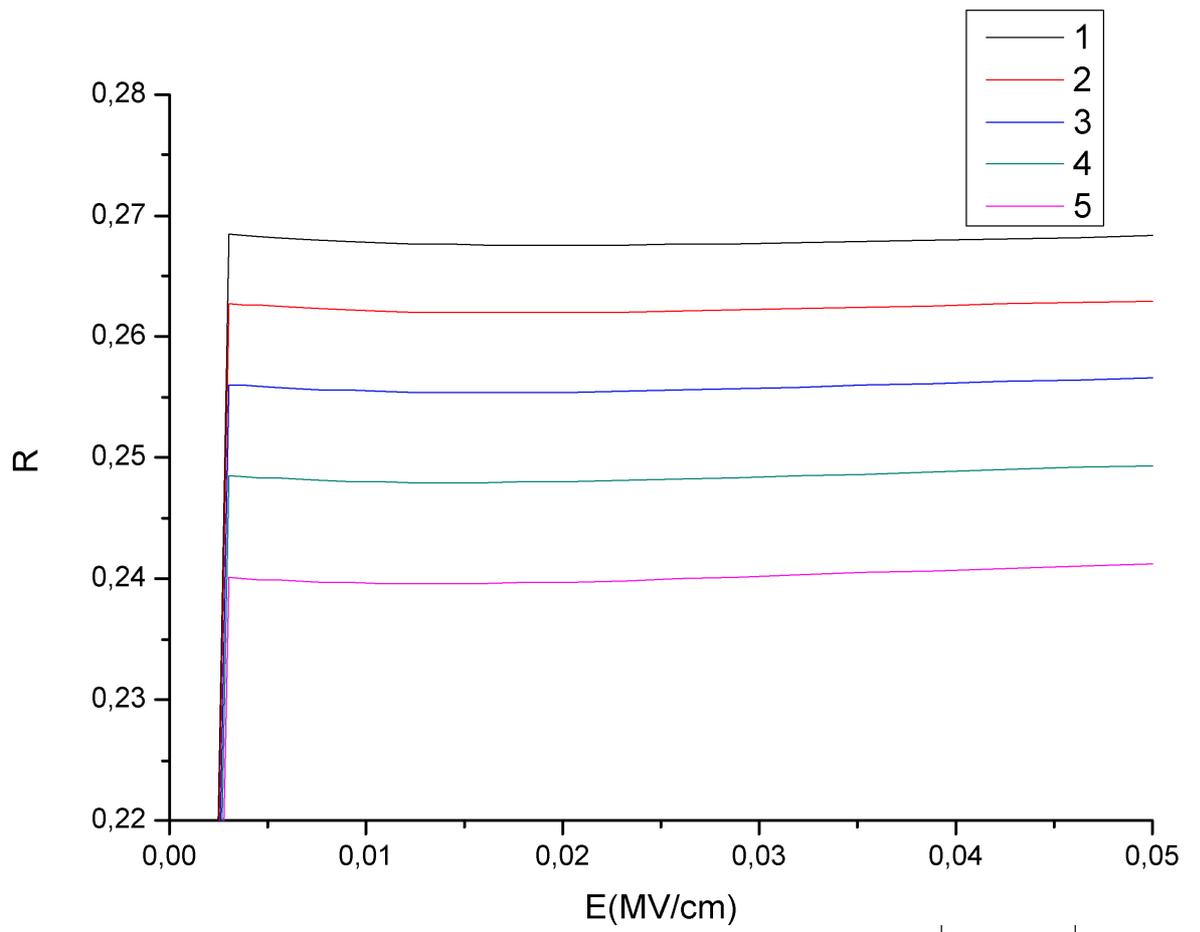

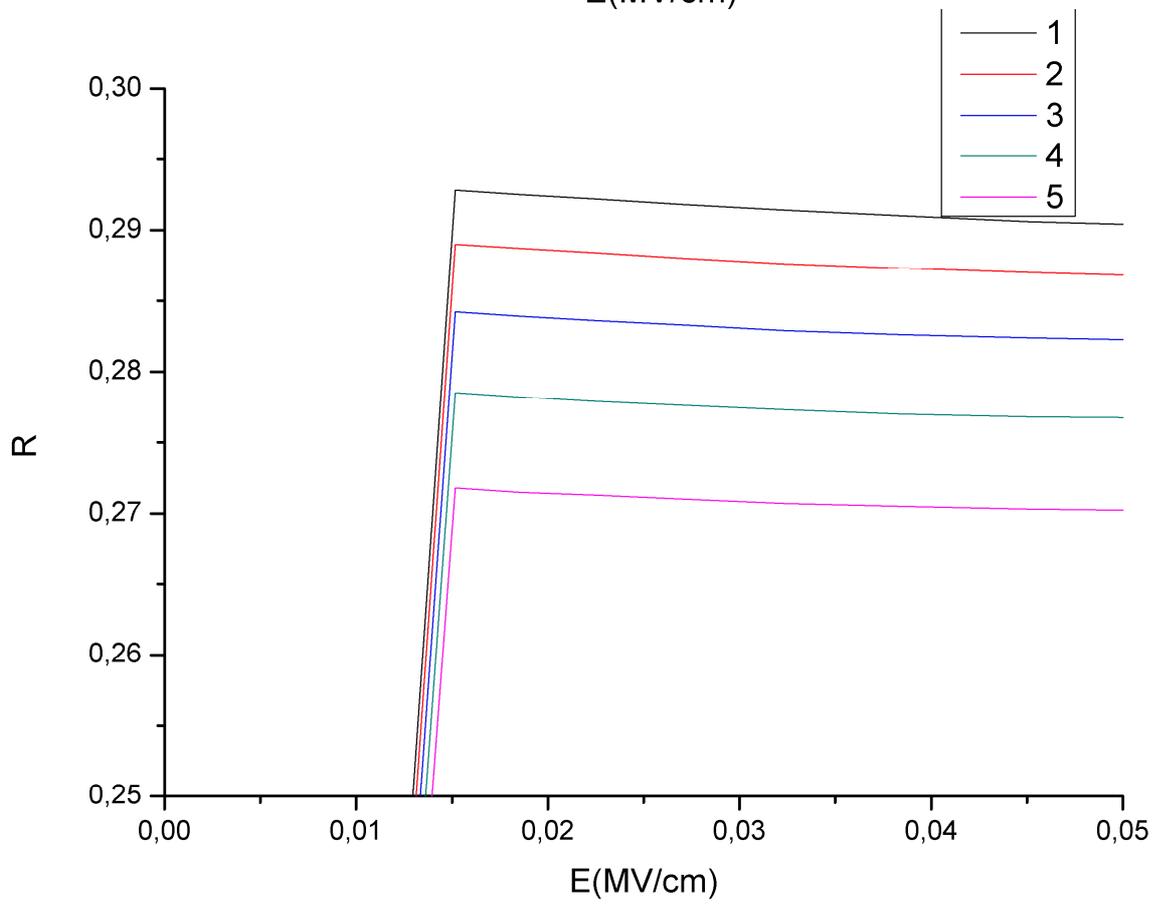

Fig.3. Reflectivity *R* at wavelength 10.6 μm dependence on in-substrate field at room temperature for different substrate thickness (curve 1 – 240 nm, 2 – 280 nm, 3 – 320 nm, 4 – 360 nm, 5 – 400 nm). A – single layer graphene, B – 5 layer graphene.

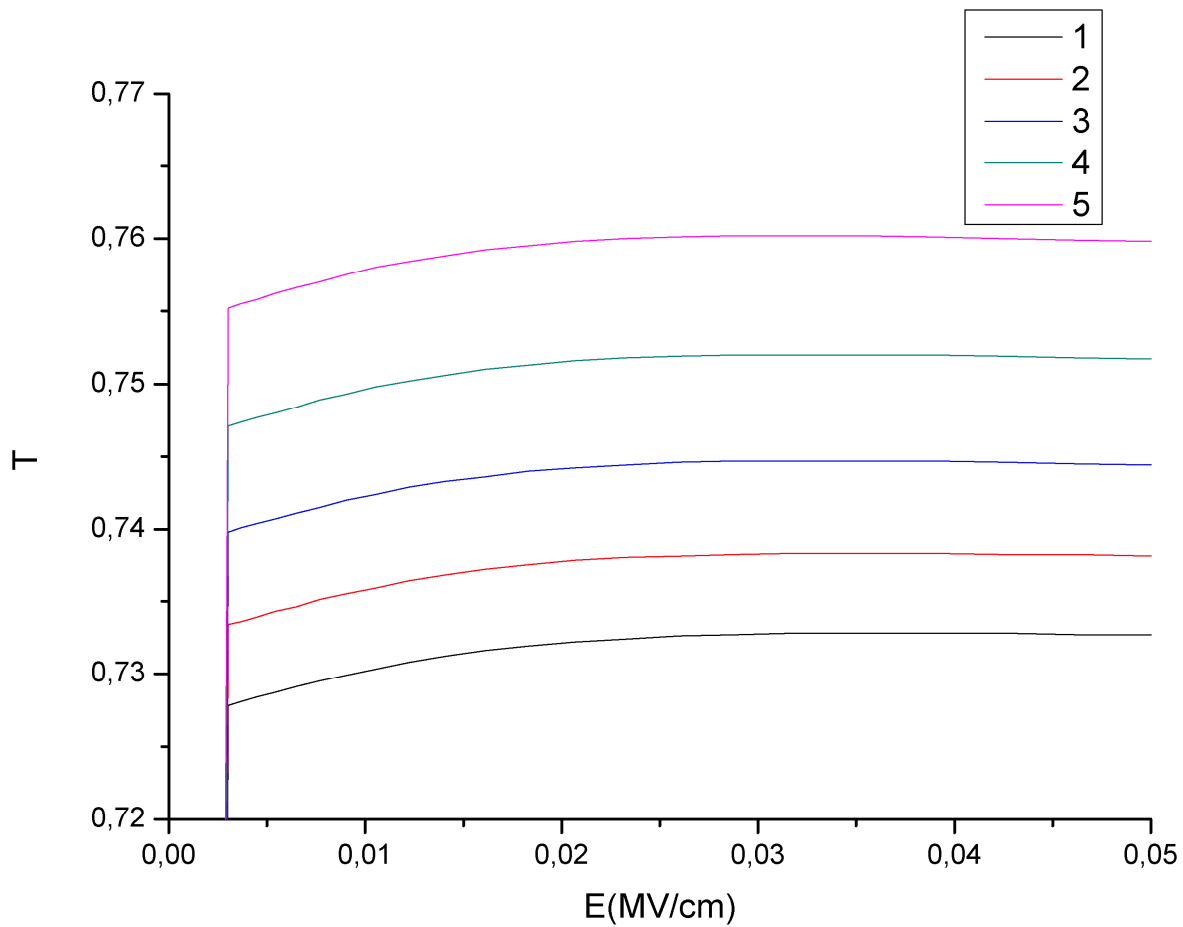

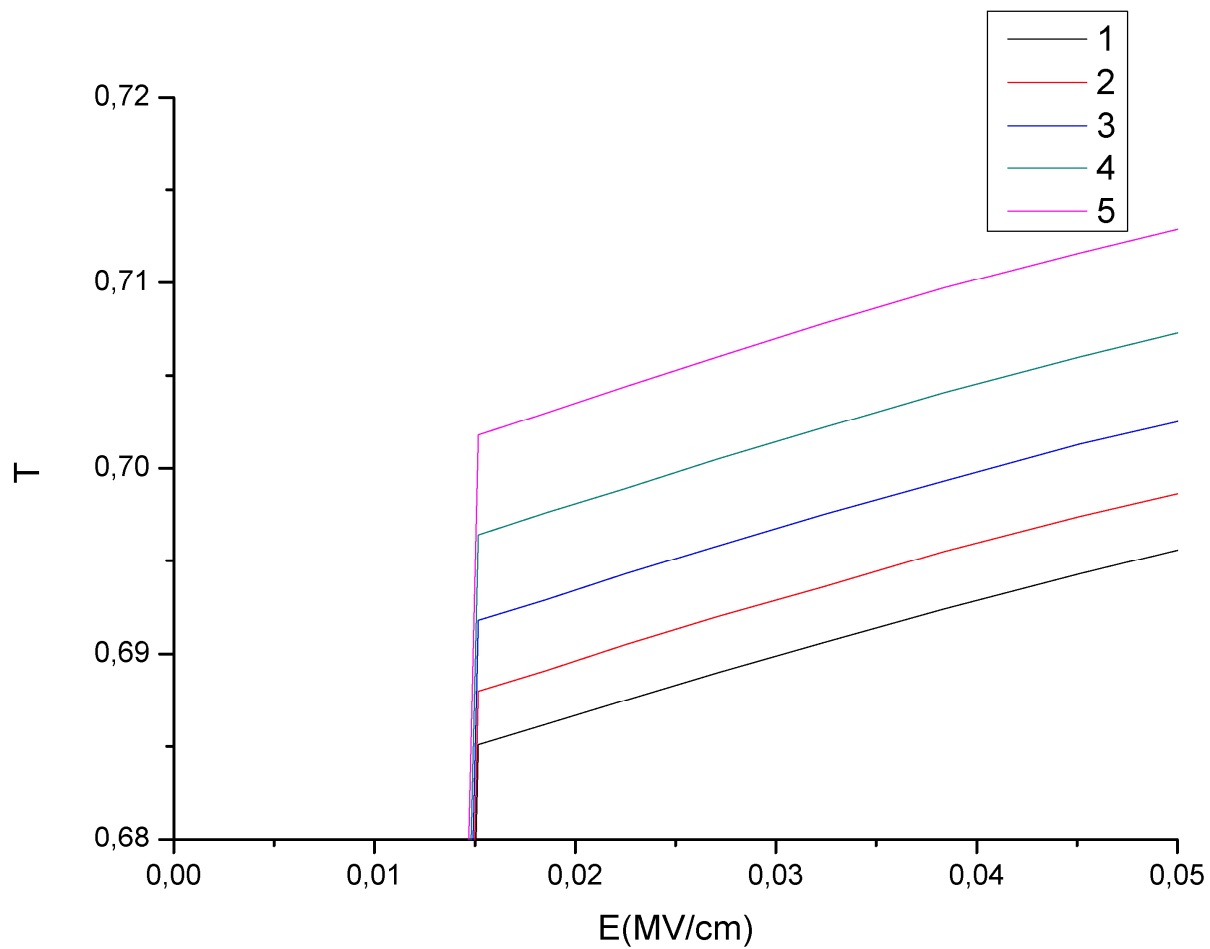

Fig.4. Transmittivity *T* at wavelength 10.6 μm dependence on in-substrate field at room temperature for different substrate thickness (curve 1 – 240 nm, 2 – 280 nm, 3 – 320 nm, 4 – 360 nm, 5 – 400 nm). A – single layer graphene, B – 5 layer graphene.